\documentstyle[preprint,tighten,aps,prd,amssymb,eqsecnum]{revtex}     
\begin{document}
\draft
\preprint{DAMTP-R96/59}
\title{The tale of two centres}
\author{ Neil J. Cornish and Gary W. Gibbons} 
\address{DAMTP. University of Cambridge, Silver Street, Cambridge CB3
9EW, UK}
\maketitle
\begin{abstract}
We study motion in the field of two fixed centres described by a
family of Einstein-dilaton-Maxwell theories. Transitions between
regular and chaotic motion are observed as the dilaton coupling is
varied.

\end{abstract}
\pacs{05.45.+b, 04.20.-q, 04.50, 04.65, 11.80}


\setcounter{section}{1}

The classical two centre problem describes the motion of a small
mass in the field of two fixed centres. The solution for motion
restricted to a plane was given by Euler in 1760, and the general
solution was found by Jacobi\cite{jac} in 1842. In the Newtonian case
the centres may be kept fixed by
balancing gravitational attraction against electrostatic
repulsion. The relativistic analog of the two centre geometry was found
independently by Majumdar\cite{maj} and Papapetrou\cite{pap}, and was
later shown\cite{hawky} to describe two or more extremal
Reissner-Nordstrom black holes. The general relativistic two (or $N$) centre
spacetime has now been extended to form a one-parameter family of
solutions\cite{gary}. These solutions contain
a dilaton field $\phi$ in addition to the gravitational field $g$ and Maxwell
field $F=dA$ and are described by the action
\begin{equation}
S=\int d^4x\, \sqrt{-g}\, \left( R-2(\partial \phi )^2
-e^{-2a \phi}F^2 \right) \, .
\end{equation}
The static $N$-centre solution solution takes the form
\begin{eqnarray}\label{two}
&&g=-H^{-2/(1+a^2)}dt^2+H^{2/(1+a^2)}d{\bf x}\cdot d{\bf x}\, ,
\nonumber \\
&& A={1 \over \sqrt{1+a^2}H}\, dt \, , \nonumber \\
&& e^{-\phi} = H^{a/(1+a^2)} \, ,
\end{eqnarray}
where $H$ is a harmonic function describing the positions
${\bf x}_{i}$ of the masses $M_i$:
\begin{equation}
H=1+\sum_{i} { (1+a^2) M_{i} \over |{\bf x}-{\bf x}_{i}| } \, .
\end{equation}
Each black hole has mass $M_i$, electric charge $Q_i$ and dilaton
charge $\Sigma_i$. These charges satisfy the extremality condition
$Q^2=M^2+\Sigma^2$. The parameter $a$ labels the family of
solutions, and is related to various reductions of $N=8$ supergravity.
Some interesting special cases are: $a=0$ Einstein-Maxwell;
$a=1/\sqrt{3}$ Einstein-Maxwell reduced from $5\rightarrow 4$
dimensions; $a=1$ string theory; $a=\sqrt{3}$ Einstein gravity reduced
from $5\rightarrow 4$ dimensions . Solutions with
$a=0,1/\sqrt{3},1,\sqrt{3}$ arise in type II
string theory as marginally bound states of elementary
solutions with $a=\sqrt{3}$. Much of the recent interest in these
solutions has focused on the duality between extremal black holes and
intersecting D-branes\cite{craze}. This broader context is not the
focus of our current study. Rather, we are interested in using
(\ref{two}) to study the relativistic two centre problem.

The Newtonian two centre problem leads
to equations of motion that are separable and hence
integrable\cite{jac}. Chandrasekhar\cite{chandra} considered the
Einstein-Maxwell two centre problem but was unable to integrate the
equations of motion. Methods borrowed from dynamical systems theory
were then used to prove that null and timelike geodesics of the
Majumdar-Papapetrou spacetime were chaotic\cite{cod,dfc}. Here we
extend these results to include null and timelike geodesics of the general
Einstein-Maxwell-dilaton two centre spacetimes. Most of the
paper is devoted to  null geodesics as the chaotic dynamics admit a
complete, analytic description. The null
geodesics can be thought of as describing ultra-relativistic chaotic
scattering. Timelike geodesics and the motion of extremally charged test
particles are discussed at the end of the paper. Extremal test
particles have charges that satisfy
$q^2=m^2+\sigma^2$, where $m$ is the mass, $q$ is the electric charge
and $\sigma$ is the dilaton charge.
The motion of an extremally charged test particle in the
field of two massive fixed centres provides a first approximation to motion
in moduli space - the space of all static solutions.

It  is well known that null geodesics in static spacetimes correspond to 
ordinary geodesics of a  three dimensional optical
metric which, in our case, is given by
\begin{equation}
ds^2_o = H^ { 4 /(1+ a^2) } d {\bf x} ^2
\end{equation}
The optical metric is defined on ${\Bbb R}^3/ \{{\bf x}_i\}$
and it is  complete if $a^2 \le 1$. If $a^2<1$ we have
an outer asymptotically flat region connected to a number of asymptotically
flat ($a=0$) or asymptotically conical ($0<a<1)$ regions surrounding
the centres.
The asymptotic regions are separated by Einstein-Rosen type \lq throats\rq.
If $a^2=1$ these throats become infinitely long tubes. If $a^2 >1$ then
the optical metric is incomplete and the centres are singularities
of the optical metric at finite optical distance.

Rather remarkably, we
find that null geodesics in the Kaluza-Klein limit $a=\sqrt{3}$ are
non-chaotic, and can be integrated using methods familiar to
Jacobi\cite{jac}. For intermediate values of $a$ we find that
the chaotic bands in phase space grow as $a$ increases from
$a=0$, reaching a peak at the string value $a=1$, and then shrinking
to zero as $a\rightarrow \sqrt{3}$.

The story for uncharged $(q=\sigma=0)$ timelike geodesics
is not so rich as chaos reigns for all values of $a$. In contrast, the
character of the dynamics for extremal test particles varies strongly
with $a$. Extremal test particles behave very much like photons. A
moving extremal particle never comes to rest. When moving, extremal
particles interact via purely velocity dependent forces -- there is no
static component to the attraction. The transition pattern between
regular and chaotic motion mirrors that of null geodesics. In
particular, we find that the motion of a small extremal black hole in
the field of two larger black holes is integrable in the Kaluza-Klein limit.

\section{Regular and chaotic motion}

\subsection{Time and chaos}

\begin{quote}
It was the best of times, it was the worst of times.\\
\hspace*{0.9in} {\it Dickens, The tale of two cities}
\end{quote}

When studying the dynamics of generally covariant theories, one no
longer has the rigid Newtonian concepts of absolute space and universal
time. Time and space become relative concepts, and standard measures
of chaos that rely on the metrical properties of phase space become
observer dependent. In our ``tale of two centres'' there is no choice of
time coordinate that is any better or any worse than any other.

For this reason, we abandon gauge dependent measures such as Lyapunov
exponents and metric entropy in favour of invariant measures such as
fractal dimensions and topological entropies.

\subsection{Fractal methods}

The geodesics will be chaotic if there exists a {\em chaotic
invariant set of orbits}. This set is usually referred to as a strange
repellor or strange saddle. The terminology ``strange'' refers to the
set having some non-integer multi-fractal dimensions. The term
``repellor'' or ``saddle'' indicates that the orbits are unstable in
some directions and stable in others. Uncovering these fractal
structures in phase space provides a gauge invariant way of showing that
the motion is chaotic\cite{dfc,nj}

Of particular interest is the future
invariant set. For unbound motion, the future invariant set
correspond to those trajectories
that approach the two centres from infinity with an impact parameter
that allows them to take up everlasting periodic orbits.
Since these orbits never exit the scattering region, they
cannot be assigned to a particular asymptotic outcome. Thus, the
future
invariant set forms the boundary between the various 
outcomes.
In a numerical experiment the future invariant set can be
uncovered by
studying the nature of the boundaries between the outcome basins.

The experimental setup is shown schematically in Fig.~1. The two
centres, both with mass $M=1$, are placed at ${\bf x}=(0,\pm
1,0)$. Since angular momentum is conserved about the $y$-axis, it is
enough to consider motion restricted to the plane $z=0$. Null
geodesics are fired into the scattering region from the point
$(x_0,y_0)$ at an angle $\theta_0$ from the $x$-axis. The initial
velocities are given by
\begin{equation}
u^x={\cos \theta_{0} \over H(x_0,y_0)^{2(1+a^2)}}\, \quad
u^y={\sin \theta_{0} \over H(x_0,y_0)^{2(1+a^2)}}\, .
\end{equation}

The trajectories are evolved numerically until an outcome is
reached. We assign four different outcomes on the basis of where the
trajectory ends up in the $(x,y)$ plane:
\begin{equation}
{\rm I}\; (0,1), \;\; {\rm II}\; (0,-1),\;\; {\rm III}\; (\pm\infty , \infty
), \; \; {\rm IV}\; (\pm \infty, -\infty ) \, .
\end{equation}
The first two of these outcomes will not occur for all values of $a$,
since the capture cross section of each centre goes to zero for $a>1$.
Initial conditions
were chosen by setting $x_0=10$ and selecting $(y_0,\theta_0)$ from a
$500^2$ grid. Points in this grid are colour coded according to their
final outcome using the colour scheme: I dark grey; II light grey; III
black; IV white. The results of the numerical experiment are displayed
in Fig.~2 for three values of $a$. The first two graphs show evidence
of ``chaotic gravitational lensing''\cite{janid}.

The boundaries between the outcomes appear to be fractal for $a=0,1$ and
smooth for $a=\sqrt{3}$. To confirm these impressions, portions of
each image are enlarged in Fig.~3. Again, the boundaries for $a=0,1$
are clearly fractal, while the boundaries
for $a=\sqrt{3}$ look quite smooth.

As a further check, the capacity or box
counting dimension 
of the images is computed using the method described in
Ref.~\cite{dfc}. We find
\begin{eqnarray}
&& D(a=0)=1.36 \pm 0.02 \, ,\nonumber \\
&& D(a=1)=1.50 \pm 0.02 \, , \nonumber \\
&& D(a=\sqrt{3})=1.01 \pm 0.02 \; .
\end{eqnarray}
Within numerical tolerances, these dimensions support the contention
that null geodesics in geometries with $a=0,1$ are chaotic, while the
$a=\sqrt{3}$ geometry has regular null geodesics. Repeating this
analysis for several values of $a$ we find that geometries with
$a<\sqrt{3}$ display chaotic scattering, while those with $a\geq
\sqrt{3}$ do not. The chaotic behaviour is most pronounced at the
string value $a=1$.

\subsection{Curvature methods}

As an alternative to our fractal methods, several
groups~\cite{syd,yurt,rugh,maeda} have advocated curvature as a
coordinate independent tool for forecasting chaos. The idea is to
extend Hadamard's~\cite{hat} classic result that the geodesic flow on
a compact manifold with all sectional curvatures negative at every
point is chaotic. So far, these attempts have failed to yield any
reliable method to forecast chaos. The various criteria proposed are
neither necessary nor sufficient for predicting chaos. 

Despite these shortcomings, the idea of using curvature methods is not
entirely without merit. With a little care it is possible to arrive at
a necessary, but not sufficient, criteria for chaos. The reasoning is as
follows: For chaos to occur the phase space must contain a chaotic
invariant set. Since the elements of this set are unstable periodic
orbits, the dynamics must admit such trajectories. If there are no
unstable periodic orbits, then the dynamics will not be chaotic.
Applying this test to geodesic motion requires us to show (1) the
manifold admits periodic geodesics, and (2), most of these orbits are
unstable. A sufficient criteria for condition (2) to hold can be given
in terms of orbit-averaged sectional curvatures.
Chaotic behaviour is ruled out if either (1) or (2) is not satisfied.

To improve our test so that it is both a necessary and sufficient for
chaos would require some notion of mixing. For Hadamard, the mixing
comes about
because the hyperbolic manifold is compact. In general, the geodesic
flow will not be restricted to a compact region, so we have to look
for other mixing mechanisms. For near-integrable systems the mixing
can be caused by a homoclinic or hetroclinic tangle, the existence of
which can be probed using the Melnikov method\cite{mel}.

Condition (1), i.e.~the existence of periodic null geodesics, is
equivalent in a static spacetime to the existence of closed geodesics
in the optical metric. Some partial information is provided by 
the Benci-Giannoni theorem\cite{bg}. The Benci-Giannoni theorem guarantees
the existence of at least one non-constant closed geodesic of a complete
(but not necessarily compact) riemannian manifold provided a
condition on the fall-off of the sectional curvatures
 and a topological condition hold. The  optical metric 
is complete if $a\leq 1$ and the topological condition
holds in that case. The sectional curvature condition
for the optical metric
is easily seen to hold if $a<1$. Thus,
perhaps surprizingly,  the Benci-Giannoni theorem
guarantees the existence of at least one, and probably
very many, exactly periodic null geodesics no matter how 
many centres one has and no matter how one positions them.

Condition (2) can be checked using the geodesic deviation equation.
Returning to the usual spacetime metric,
the geodesic deviation equation describes how nearby geodesics
separate:
\begin{equation}\label{geodev}
{D^2 n^\mu \over D s^2}=-R^{\mu}_{\;\, \nu\rho\sigma}u^{\nu}
n^{\rho}u^{\sigma} \, .
\end{equation}
We demand that the deviation $n^{\mu}$ is a spacelike four vector
orthogonal to the four velocity $u^{\mu}$, {\it ie.} 
\begin{equation}
n^{\mu}u_{\mu}=0 \quad u^{\mu}u_{\mu}=0,\, -1 \quad n^{\mu}n_{\mu}>0.
\end{equation}
It is convenient to introduce a set of non-rotating orthonormal basis vectors,
$E^{\alpha}_{a}$ $(a=0..3)$, with $E^{\alpha}_0$ set equal to
$u^{\alpha}$\cite{hawkellis}. The remaining three basis vectors can
then be used to describe the spacelike deviation vector
$n^{\mu}=n^{i}E_{i}^{\mu}$ where $i=1..3$. Only two basis vectors are
required to describe the deviation when the four velocity is
null\cite{hawkellis}. The advantage
of this approach is that it separates the rotation of the geodesic
congruence from its spreading. This allows us to write the deviation equation
(\ref{geodev}) in terms of ordinary, rather than covariant derivatives:
\begin{equation}\label{geodev2}
{d^2 n^{i} \over ds^2}=-R^{i}_{\;\, 0j0}n^{j}\, .
\end{equation}
Contracting (\ref{geodev2}) with $n_i$ and averaging over an orbit we find
\begin{equation}
{\triangle 
(dn^2/ds) \over <\! n^2 \! >}=-<\! K({\bf u},{\bf n})\! >+{<\!
(dn^i/ds)^2\! > \over <\! n^2\! >}, \label{sec}
\end{equation}
where $n=(n^\mu n_\mu)^{1/2}$,
\begin{equation}
<\! K({\bf u},{\bf n})\! > \, \equiv { <\! R_{\mu\nu\kappa\lambda}
n^{\mu}u^{\nu}n^{\kappa}u^{\lambda}\! >
\over <\! n^{\mu}n_{\mu} \! >} \, ,
\end{equation}
and
\begin{equation}
<\! f \! >=\oint f\, ds=\int_{0}^{0'}\! f \, ds\, ,\quad \triangle f =
f\Big\vert_{0}^{0'} \, .
\end{equation}
The quantity $<\! K({\bf u},{\bf n})\! >$ is essentially the sectional
curvature in the plane spanned by $u^{\mu}$ and $n^{\mu}$, averaged
over one orbit. If $<\! K({\bf u},{\bf n})\! >$ is negative we have
\begin{equation}
\triangle  (dn^2/ds) > 0 \, .
\end{equation}
That is, the rate of deviation increases with each orbit:

{\em A periodic orbit is unstable if any of its three (two for null geodesics)
average sectional curvatures is negative.}

To see that conditions (1) and (2) only provide a necessary condition,
we can apply the test to a single black hole spacetime with $a=0$. In
this case there is a circular photon orbit at $r=2M$ (in areal rather
than isotropic coordinates) with four velocity
\begin{equation}
u^{\mu}=(1,0,0,{ {\scriptstyle 1} \over {\scriptstyle 4M}}) \, .
\end{equation}
A suitable pseudo-orthonormal set of basis vectors are
\begin{eqnarray}
&E_0^{\mu}=u^{\mu}, \quad & E_3^{\mu}=(4,0,\frac{ {\scriptstyle 1} }{
{\scriptstyle M}},0) \nonumber \\
&E_1^{\mu}=(0,{{\scriptstyle1}\over {\scriptstyle 2}},0,0), \quad & 
E_2^{\mu}=(0,0,{ {\scriptstyle 1} \over
{\scriptstyle 2M}},0) \, .
\end{eqnarray}
Both ${\bf E}_0$ and ${\bf E}_{3}$ are null and satisfy
$g_{\mu\nu}E_0^{\mu}E_3^{\nu}=-1$. Since we are dealing with null
geodesics, we need only consider deviations in the plane spanned by
${\bf E}_1$ and ${\bf E}_2$. Writing the deviation vectors as
${\bf l}=l{\bf E}_1$ and ${\bf q}=q{\bf E}_2$ we find
\begin{equation}
<\! K({\bf u},{\bf l})\! >=-{1 \over 32M^2}\, ,\quad 
<\! K({\bf u},{\bf q}) \! >={1 \over 16M^2}\, . 
\end{equation}
The rate of contraction in the ${\bf l}$ direction exceeds the rate
of expansion in the ${\bf q}$ direction due to Ricci focusing.
This focusing is due to the term
\begin{equation}
R_{\mu\nu}u^{\mu}u^{\nu}={1 \over 32 M^2} \, ,
\end{equation}
in the Raychaudhuri equation for the expansion,
$\theta=u^{\mu}_{;\mu}$. Despite the focusing term, the circular
photon orbit is unstable against radial perturbations. Thus, a single
extremal Reissner-Nordstrom black hole spacetime satisfies the
necessary conditions for chaos to occur. However,
geodesics are integrable in this spacetime, so our
curvature condition is not a sufficient condition for chaos.

Returning to the geodesic equation for $l^{\mu}$, we have
\begin{equation}
{d^2 l \over d t^2}={1 \over 32 M^2}\, l \, .
\end{equation}
Here we have chosen the affine parameter for the null geodesic to
coincide with the coordinate time $t$.
Solving for $l(t)$ we find
\begin{equation}
l=l_0 \exp\left({t\over \sqrt{32}\, M}\right) \, .
\end{equation}
One might be tempted to say that the radial direction has a positive
Lyapunov exponent $\lambda=1/(\sqrt{32}\, M)$, but this statement is
highly coordinate dependent. For example, an observer free falling
into the black hole sees the trajectories diverge at a rate
$l(\tau)\sim \tau^{2\sqrt{2}}$, so she would conclude that
$\lambda=0$. To avoid this type of ambiguity, we define unstable to
mean there is at least one negative orbit-averaged sectional curvature.

\section{Symbolic Dynamics}

Taken on their own, the numerical results provide a solid, but
rather unenlightening description of the dynamics. A far more
satisfying description can be found using symbolic dynamics. Symbolic
dynamics describes the topology of trajectories in phase
space. Because the detailed local dynamics does not enter into this
description, the symbolic dynamics can be studied analytically, even
when the trajectories are not integrable.

Since we are considering chaotic scattering, the unstable periodic
orbits that form the strange repellor are of particular
importance. Typical scattering trajectories experience chaotic
transients as they pass through the scattering region. The dynamics of
these transients is completely encoded by the strange
repellor. Thus, by uncovering the symbolic coding of the repellor, we
learn a great deal about the dynamics.

To find the coding we place windows in phase space, positioned
in such a way that topologically distinct periodic orbits pass through
the windows in a distinct order. Each time an orbit passes through a
window, the symbol for that window is recorded. The resulting string
of symbols is the symbolic coding of the orbit. A unique symbolic
coding can be found by demanding that each distinct physical orbit has
a distinct symbolic coding, and that every symbolic coding
describes a physical orbit.

Once the coding has been found, the symbolic complexity of the
repellor can be measured. If the dynamics is chaotic, then the number
of periodic orbits will grow exponentially as the symbolic length of
the orbits is increased.

\subsection{Allowed Orbits}
The primary closed null geodesics of the two centre spacetime are
shown in Fig.~4. Orbits of type a) and b) require that
each centre is capable of bending a trajectory through at
least $\Delta_{b}=2\pi$. Similarly, type c) requires $\Delta_{b}\geq
\pi$, and type d) requires $\Delta_b > \pi$.

Since large angle scattering occurs for trajectories with small impact
parameter $b$, {\it ie.} close to one of the two centres, the maximum
scattering angle produced by each centre can be approximated by
ignoring the distant centre and considering a
one-centre spacetime. The spherical symmetry of the one-centre
spacetime reduces the dynamics to one dimensional motion in an
effective potential $V(r)$:
\begin{equation}
\left({d r \over d t   }\right)^2 = {1 \over b^2}-V^{2}(r) \, ,
\end{equation}
where $b$ is the impact parameter and
\begin{equation}
V(r)=\left(1+{(1+a^2)M \over r}\right)^{-2/(1+a^2)}r^{-1} \, .
\end{equation}
The scattering angles can be calculated from the equation
\begin{equation}
\left({ d r \over d \theta}\right)^2=r^2\left({1 \over b^2
V^2}-1\right)\, .
\end{equation}
The result is
\begin{equation}\label{inter}
\Delta^{a}_{b}=\int_{0}^{u_{{\rm max}}} {2 b\, du \over
\sqrt{(1+(1+a^2)Mu)^{a/(1+a^2)} -u^2b^2}} \; - \pi \, .
\end{equation}
Here $u=1/r$ and $u_{{\rm max}}$ denotes the point of closest
approach. The integral (\ref{inter}) will be finite unless the denominator
admits a double root. A double root occurs when
\begin{equation}
2M(1+(1+a^2)M u)^{(1-a^2)/(1+a^2)}=b \, ,
\end{equation}
and
\begin{equation}
(1+(1+a^2)Mu)^{2/(1+a^2)}=ub \, .
\end{equation}
Solving for $u$ and $b$ we find
\begin{equation}
u_{{\rm crit}}={ 1 \over M}\left({1 \over 1-a^2}\right) \, ,
\end{equation}
and
\begin{equation}
b_{{\rm crit}}=2M \left({2 \over
1-a^2}\right)^{(1-a^2)/(1+a^2)} \, .
\end{equation}
It is no coincidence that these are the same values of $u$ and $b$ for
which give rise to unstable photon orbits. We see that unstable photon
orbits are only possible if $a<1$. Thus, the scattering angle
$\Delta^a_b$ is finite for all $a>1$. The limiting case $a=1$ has
to be handled separately. A direct evaluation yields
\begin{equation}
\Delta^{1}_{b}=\pi \left[ {b \over \sqrt{b^2-4M^2}}-1\right] +
{2 b \over \sqrt{b^2-4M^2}} \sin^{-1}\left({2M \over b}\right) \, .
\end{equation}
Here the critical impact parameter separating capture from scattering
is $b_{{\rm crit}}=2M$. If we write $b=2M(1+\epsilon^2)$ where
$\epsilon\ll 1$, then we find
\begin{equation}
\Delta^1_{b}\simeq {2\pi \over \epsilon} -\pi - 4 \, .
\end{equation}
Hence the scattering angle can be infinite when $a=1$. This means a glory is
possible even though the effective potential does not have a turning
point.

For all $a>1$ the scattering angle will be finite. For $a$ just a
little larger than 1 the scattering angle can still be large enough
to allow several temporary photon orbits. This partial glory adds to
the complexity of the allowed periodic orbits, but not to the same
degree as a full glory.

 Unstable photon orbits will still be possible
in the two centre spacetimes so long as $\Delta_b > \pi$. Note that
the largest scattering angles occurs for orbits that pass very
close to one of the centres. For these orbits our approximation is
especially good. We find that the
critical value of $\Delta_b$ is reached when $a=\sqrt{3}$.
For small $b$ the scattering angle reaches
\begin{equation}
\Delta^{\sqrt{3}}_{b}= \pi - {b \over M} 
+{ b^3 \over 12 M^3}-\dots \, .
\end{equation}
In this case there are can be just one unstable periodic orbit in the two
centre spacetime, and thus no chaos. Later we will show that the
Kaluza-Klein limit $a=\sqrt{3}$ is actually integrable for null geodesics.

The existence of periodic null geodesics and the associated glories
in these black hole spacetimes should be constrasted with their complete
absence in cosmic string spacetimes \cite{garyglory}. The reason for
this difference is that for cosmic strings the sectional curvatures
are non-negative \cite{garyglory}. This fact again points to the 
importance of the sectional curvatures 
in general relativity as a 
possible diagnostic tool for chaos or its absence. 

\subsection{The symbolic coding}

Using the primary unstable orbits as a guide, we see that it is
natural to place three windows along the axis connecting the two
centres. The placement and labelling of the windows is shown on the
left of Fig.~5. Using these windows, all orbits can be represented by
strings of 0's and $\pm1$'s, with the
restriction that no symbol follows itself. The allowed transitions are
shown schematically on the right of Fig.~5.

Since a complete orbit
must contain an even number of symbols, the counting of
orbits is made easier if we shift to the new symbolic alphabet
$A=\{(1,-1),(-1,1)\}$, $B=\{(0,-1),(-1,0)\}$, $C=\{(0,1),(1,0)\}$.
We define the length of a symbolic sequence to be the sum of the
number of incontractible loops around each centre. This number is a
topological invariant as motion is restricted to the $(x,y)$ plane.
The labelling of the windows was chosen so that the length of an orbit
is given by the sum of the absolute values of the symbols used to
describe the orbit. For example, the primary orbit d) has the symbolic
coding $\overline{1,0,-1,0}=\overline{CB}$, and is thus of length
$k=2$. Here overline denotes a sequence to be repeated. 

The counting of orbits as a function of their length is a simple
exercise in combinatorics. An orbit of length $k=2p+n$ will be made up
of $n+p$ symbols \{$A,B,C$\},
where $p$ is the number of $A$'s, $i$ the number of $B$'s and $n-i$ the
number of $C$'s. Thus, the number of orbits at order $k$ is given by
\begin{eqnarray}\label{nk}
&&N(k)=\sum_{p=0}^{[k/2]}\; \sum_{i=0}^{n=k-2p} { (n+p)! \over p!\, i!\,
(n-i)!} \nonumber \\
&&\quad = {1 \over 2^{3/2}}\left( (\sqrt{2}+1)^{k+1}+(-1)^k {1 \over
(\sqrt{2}+1)^{k+1}} \right) \, .
\end{eqnarray}
Here the notation $[x]$ denotes the integer part and the double sum
is over a trinomial combinatoric factor.
Readers familiar with number theory will recognise
$\sqrt{2}+1$ as the silver mean $2+1/(2+1/(2+\dots$. The expression
for $N(k)$ looks more natural when derived from a recurrence
relation. Longer orbits can be generated from shorter orbits by
inserting $A$'s, $B$'s and $C$'s, so that
\begin{equation}\label{rr}
N(k+2)=N(k)+2 N(k+1) \, .
\end{equation}
The first term comes from inserting an $A$, and the second from
inserting a $B$ or a $C$. A direct counting of orbits reveals $N(1)=2$
and $N(2)=5$. Using these initial values, one immediately arrives at
(\ref{nk}) as the solution to the recurrence relation (\ref{rr}).

\subsection{Topological entropy}

Since the symbolic coding is based on an uneven three symbol coding
(the $A$'s are twice as long as the $B$'s and $C$'s), we expect the
topological entropy to lie between that of a straight two symbol
coding and a straight three symbol coding, {\it ie.}, $\ln 2 < H_{T} <
\ln 3 $. This is indeed the case:
\begin{equation}
H_{T}=\lim_{k\rightarrow \infty} \frac{1}{k} \ln N(k) = \ln
(\sqrt{2}+1) \, .
\end{equation}
Unlike the metric or Kolmogorov-Sinai entropy, the topological entropy
provides a coordinate independent measure of chaos in general
relativity\cite{nj,ws}. 

For $a>1$ the symbolic coding starts to get pruned as there can only
be a finite number of orbits around each centre between excursion to
the second centre. This limits the number of $B$'s or $C$'s that can be
strung together. If the maximum scattering angle about one centre is
$\Delta_{{\rm max}}$, then the longest string of the form $B^n$ or
$C^n$ is given by $n=[\Delta_{{\rm max}}/\pi]$. Once $\Delta_{{\rm
max}}$ drops below $2\pi$, no orbits of type a) or b) can be inserted between
orbits of type c) or d). The symbolic coding can then be reduced to a
binary alphabet and the topological entropy drops to $H_{T}=\ln
2$. The topological entropy then remains unchanged until $a=\sqrt{3}$, at
which point $\Delta_{{\rm max}}\leq \pi$ and orbits of type d) are no
longer possible. The symbolic coding collapses to a single letter and
the topological entropy drops to zero. When the topological entropy of
the strange repellor vanishes, the dynamical system is
non-chaotic. This analysis is in complete agreement with the numerical
results of the previous section.

\subsection{The integrable limit $a=\protect\sqrt{3}$}

We have shown that the scattering of massless particles in the
Kaluza-Klein two centre geometry is not chaotic. The repellor has
integer capacity dimension and zero topological entropy. Now we will
show that the Kaluza-Klein two centre problem is integrable for null
geodesics.

Null geodesics are best studied in the optical metric
\begin{equation}
d\sigma^2=-dt^2+H^{4/(1+a^2)}(dx^2+dy^2+dz^2) \, .
\end{equation}
Following Jacobi\cite{jac} we adopt prolate spheroidal coordinates
\begin{eqnarray}
&& x=\sinh\psi\sin\theta\cos\phi, \quad
y=\sinh\psi\sin\theta\sin\phi, \nonumber \\
&&\hspace*{0.7in} z=\cosh\psi\cos\theta\, .
\end{eqnarray}
In these coordinates, the harmonic function $H$ is given by
\begin{eqnarray}
&&H=1+W/Q, \quad Q=\sinh^2\psi+\sin^2\theta, \nonumber \\
&&W=(M_1+M_2)\cosh\psi+(M_1-M_2)\cos\theta \, .
\end{eqnarray}
The Hamilton-Jacobi equation,
\begin{equation}
g^{\alpha\beta}{\partial {\cal S} \over \partial x^{\alpha}}
{\partial {\cal S} \over \partial x^{\beta}}=0\, ,
\end{equation}
takes the form
\begin{equation}\label{hj}
\left({\partial {\cal S} \over \partial t}\right)^2-{1 \over
H^{4/(1+a^2)} \sinh^2\psi\sin^2\theta}\left({\partial {\cal S}
\over \partial \phi }\right)^2 
 -{1 \over H^{4/(1+a^2)}Q}\left(\left({\partial {\cal S} \over
\partial \psi}\right)^2+\left({\partial {\cal S} \over
\partial \theta}\right)^2\right)=0\, .
\end{equation}
Since $t$ and $\phi$ are cyclic coordinates, their canonically
conjugate momenta, $E$ and $L$, are constants of the motion. The
Hamilton-Jacobi equation will be separable, and hence integrable if
\begin{equation}
{\cal S}=-Et+L\phi+S_{\psi}(\psi)+S_{\theta}(\theta) \, .
\end{equation}
Substituting this anzatz into (\ref{hj}) we find the system only
separates if $a=\sqrt{3}$. The integrable limit $a=\sqrt{3}$ is
characterised by an addition constant of the motion $\alpha$:
\begin{equation}\label{sep1}
\left({\partial S_{\psi} \over \partial \psi}\right)^2=
\alpha -{L^2 \over \sinh^2\psi} 
+E^2(\sinh^2\psi+(M_1+M_2)\cosh\psi) \, ,
\end{equation}
and
\begin{equation}\label{sep2}
 \left({\partial S_{\theta} \over \partial \theta}\right)^2=
-\alpha-{L^2 \over \sin^2\theta}
 +E^2(\sin^2\theta+(M_1-M_2)\cosh\theta)  \, .
\end{equation}

\section{Timelike trajectories}

Test particle with mass $m$, electric
charge $q$ and dilaton charge $\sigma$ obey the
equation of motion
\begin{equation}
{d u^{\alpha} \over d \lambda}+\Gamma^{\alpha}_{\beta\gamma}
u^{\beta}u^{\gamma}={q \over m} e^{-a\phi}F^{\alpha}_{\;
\beta}u^{\beta} -{\sigma \over m}(u^{\alpha}u^{\gamma}\nabla_{\gamma}+
\nabla^{\alpha}) \phi \, .
\end{equation}
A curious situation occurs for extremal test particles initially at
rest. They are characterised by
\begin{equation}
q=\sqrt{1+a^2}\, m, \quad \sigma= a\, m \, , \quad 
u^t=H^{1/(1+a^2)} \, ,
\end{equation}
and
\begin{equation}
{d u^{i} \over d \lambda}=-H^{2/(1+a^2)}\Gamma^{i}_{tt}
+\sqrt{1+a^2}HF^{i}_{t}-a\nabla^{i}\phi \, . \label{stop}
\end{equation}
Inserting the background solution (\ref{two}) into (\ref{stop}) yields
\begin{equation}
{d u^{i} \over d \lambda}=\left({1 \over 1+a^2}-1+{a^2
\over 1+a^2}\right){\nabla^{i} H \over H}=0 \, .
\end{equation}
Initially static extremal test particles remain at rest. On reflection
this is not so surprising. An extremal test particle can be thought of
as a small extremal black hole moving in the field of other, larger
black holes. When at rest, the small black hole acts like
another centre in the static multi-black hole solution.

\subsection{Uncharged test particles}
The simplest timelike trajectories are the timelike geodesics followed
by uncharged ($q=\sigma=0$) test particles. Depending on their energy,
uncharged test particles may be confined to the region of space
near the two centres. Those that are not confined will
eventually be captured by a black hole or scattered to infinity. Using
the same techniques we applied to null geodesics, it is easy to show
that unconfined timelike geodesics are chaotic for all values of $a$
in the range $0\leq a \leq \sqrt{3}$.

For a single centre the motion is described by 
\begin{equation}
\left({ d r \over d \tau}\right)^2=E^2-V(r)^2 \, ,
\end{equation}
where the effective potential has the form
\begin{equation}
V={1 \over H^{4/(1+a^2)}}\left(H^{2/(1+a^2)}+{L^2 \over r^2}\right) \, .
\end{equation}
Here $E$ and $L$ are the test particle's conserved energy and angular momentum
per unit rest mass. Trajectories with $E>1$ are able to escape to infinity.
As we found for the null geodesics, massive particles with non-zero
angular momentum can only be captured if $a<1$. Moreover, there are no unstable
periodic orbits if $a>1$, but there are stable orbits for all $a$.

For two centres, separability of the Hamilton-Jacobi equation,
\begin{equation}
g^{\alpha\beta}{\partial {\cal S} \over \partial x^{\alpha}}
{\partial {\cal S} \over \partial x^{\beta}}=-m^2\, ,
\end{equation}
is broken in the Kaluza-Klein case ($a=\sqrt{3}$) by the term
\begin{equation}
m^2 H^{1/2} \, . \label{nonin}
\end{equation}
As the energy of the test particle is increased,
the non-integrable term (\ref{nonin}) can be neglected so that
ultra-relativistic test particles have non-chaotic geodesics.

An interesting new ingredient enters into the motion of confined test
particles. There will be a locus of points surrounding the two centres
where confined geodesics will momentarily come to rest before falling
back toward the centres. This locus of points is called the zero
velocity curve. In a two centre spacetime the zero velocity curve
makes possible a whole new range of unstable periodic orbits not
possible for scattering trajectories. The new class of unstable orbits
comprises all possible traverses between points on the zero velocity
curve. As a result, confined geodesics are very much more chaotic than
unconfined geodesics.

The highly chaotic nature of confined geodesics is dramatically
illustrated in Fig.~6. The basin structure results from treating
timelike geodesics in a Kaluza-Klein two centre spacetime as a
Hamiltonian exit system\cite{nj,ott}. Since $a>1$ and $E<1$, the
geodesics can neither escape to infinity nor fall
into a black hole. In other words, there are no asymptotic outcomes.
To remedy this problem, we introduce exits around both centres. 
A small circle was drawn around
each centre. When a trajectory exits the system an outcome is
assigned according to which centre the exit enclosed -- black for
$(0,1)$, white for $(0,-1)$. Test particles were released from rest
with initial positions in the $(x,y)$ plane taken from a $500^2$
grid. The initial grid point was colour coded according to
its outcome.

It might appear that we are forcing a square peg into a round hole by
using fractal basin boundaries in a situation where there
are no natural asymptotic outcomes. It is certainly true that a system
with bound trajectories is well suited to study by standard techniques
such as Poincar\'{e} surfaces of section. A  Poincar\'{e} section
would also provide coordinate independent, fractal information about
the dynamics, but without the need for exits to be introduced. What we
wanted to show was that fractal basin boundaries work for {\em both}
bound and unbound systems. On the other hand,  Poincar\'{e} sections
can only be used for bound orbits and this means they are of very
limited use in general relativity\cite{nj}.

\subsection{Extremal test particles}

Earlier we showed that static extremal test particles remain at
rest. The converse is also true: If an extremal test particle is in
motion, it will never come to rest. (We are neglecting
retardation effects due to the emission of gravitational, scalar or
electromagnetic radiation). In loose terms, extremal test particles
only feel velocity dependent forces.

In many ways, an extremal test particle in motion behaves like
a photon. In fact, it can be shown\cite{gwells} that extremal test
particles moving in the metric (\ref{two}) follow null geodesics of the
five dimensional metric
\begin{eqnarray}
&&g_5=H^{4a^2/3(1+a^2)}(dx^5+\sqrt{1+a^2}A_t
dt)^2 \nonumber \\
&& \quad +H^{-8a^2/3(1+a^2)}\left(dt^2
+H^{4/(1+a^2)}d{\bf x}\cdot d{\bf x}\right) \, .
\end{eqnarray}
The five dimensional Hamilton-Jacobi equation
can be reduced to the four dimensional form\cite{gwells}
\begin{equation}\label{hj2}
\left({\partial {\cal S} \over \partial x^{\alpha} }-m\sqrt{1+a^2}\,
A_{\alpha}\right)^2=-m^2 H^{-2a^2/(1+a^2)} \, .
\end{equation}

The dynamics of extremal test particles in the various two centre
geometries can be charted using the same techniques we applied to null
geodesics. A numerical survey indicated that the pattern of dynamical
behaviour exhibited by extremal test particles is essentially
identical to what we found for null geodesics.
This correspondence appeared to be
independent of velocity, so long as the particles were moving. For
example, the outcomes basins of Fig.~7 should be compared to those
displayed in Fig.~2. The morphology of the basins is identical.

Once again we see that the Kaluza-Klein geometry appears to admit
regular trajectories. Adopting prolate spheroidal coordinates, it is a
simple exercise to show that the Hamilton-Jacobi equation (\ref{hj2})
is separable when $a=\sqrt{3}$. The only extra terms in the separation
equations (\ref{sep1}) and (\ref{sep2})
are $2mE\sinh^2\psi$ and $2mE\sin^2\theta$ respectively.

\section{Conclusions}
Using a combination of fractal and topological techniques, we have
given a complete description of the Einstein-dilaton-Maxwell family of
two centre problems. Unlike their Newtonian counterpart, most
relativistic two centre problems are not integrable. There are
arguments indicating
that for more than two centres the Newtonian problem is 
not Liouville integrable \cite{symplectic} and it therefore seems likely
that in our case as well the motion will
not be integrable for more than two centres.

We did find one exceptional case where our tale of two centres came
full circle to its Euler-Jacobi antecedent: the motion of massless
particles and small extremal
black holes is integrable in the field of two fixed centres residing
in a Kaluza-Klein compactified five dimensional spacetime. Thus, we
can add the Kaluza-Klein two centre problem to our meagre collection
of integrable three body problems.

\begin{figure}[h]
\vspace*{80mm}
\includegraphics{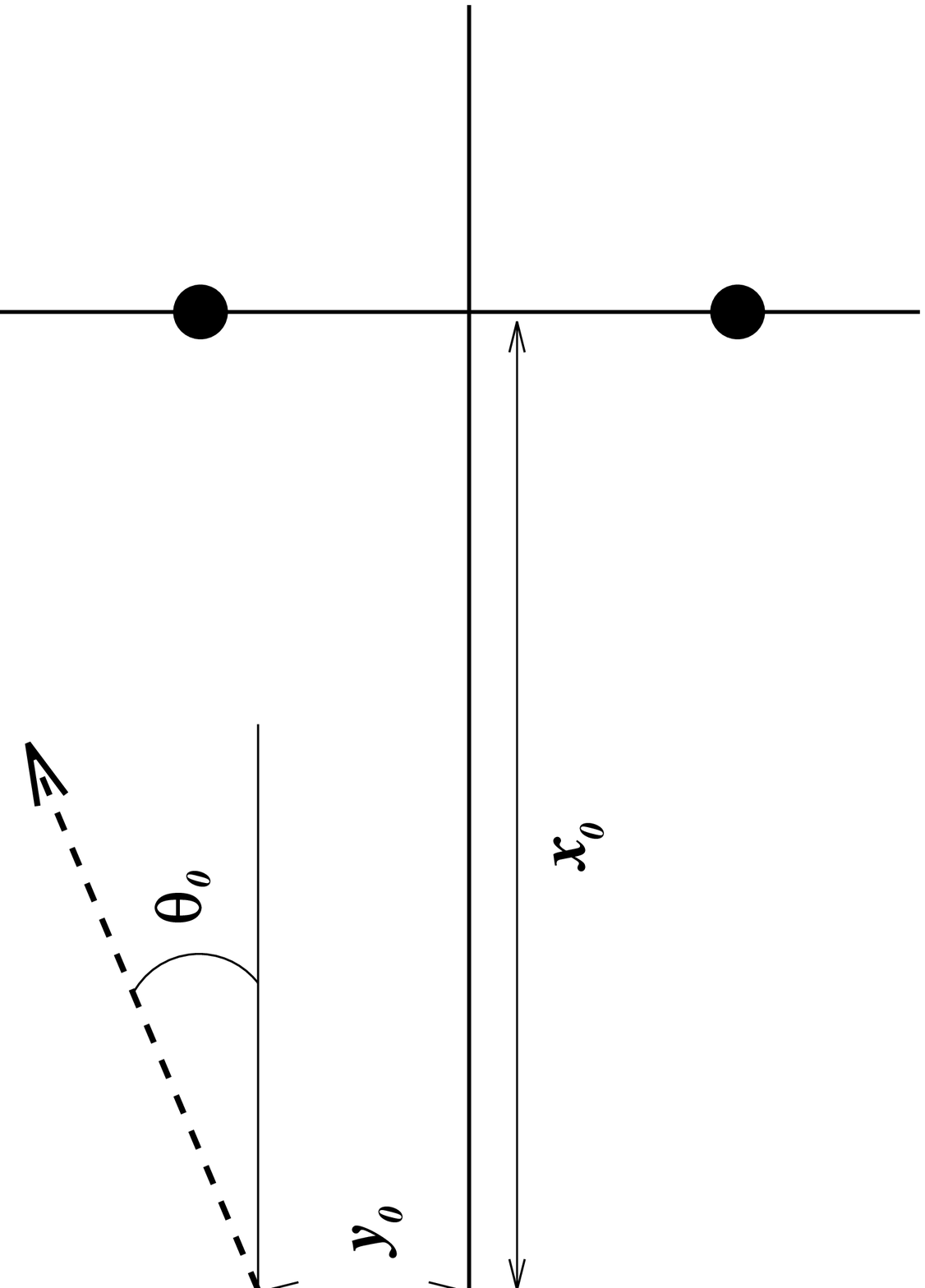}
\vspace*{20mm}
\caption{Initial conditions for the null geodesics.}
\end{figure}

\newpage

\
\begin{figure}[h]
\vspace{190mm}
\includegraphics{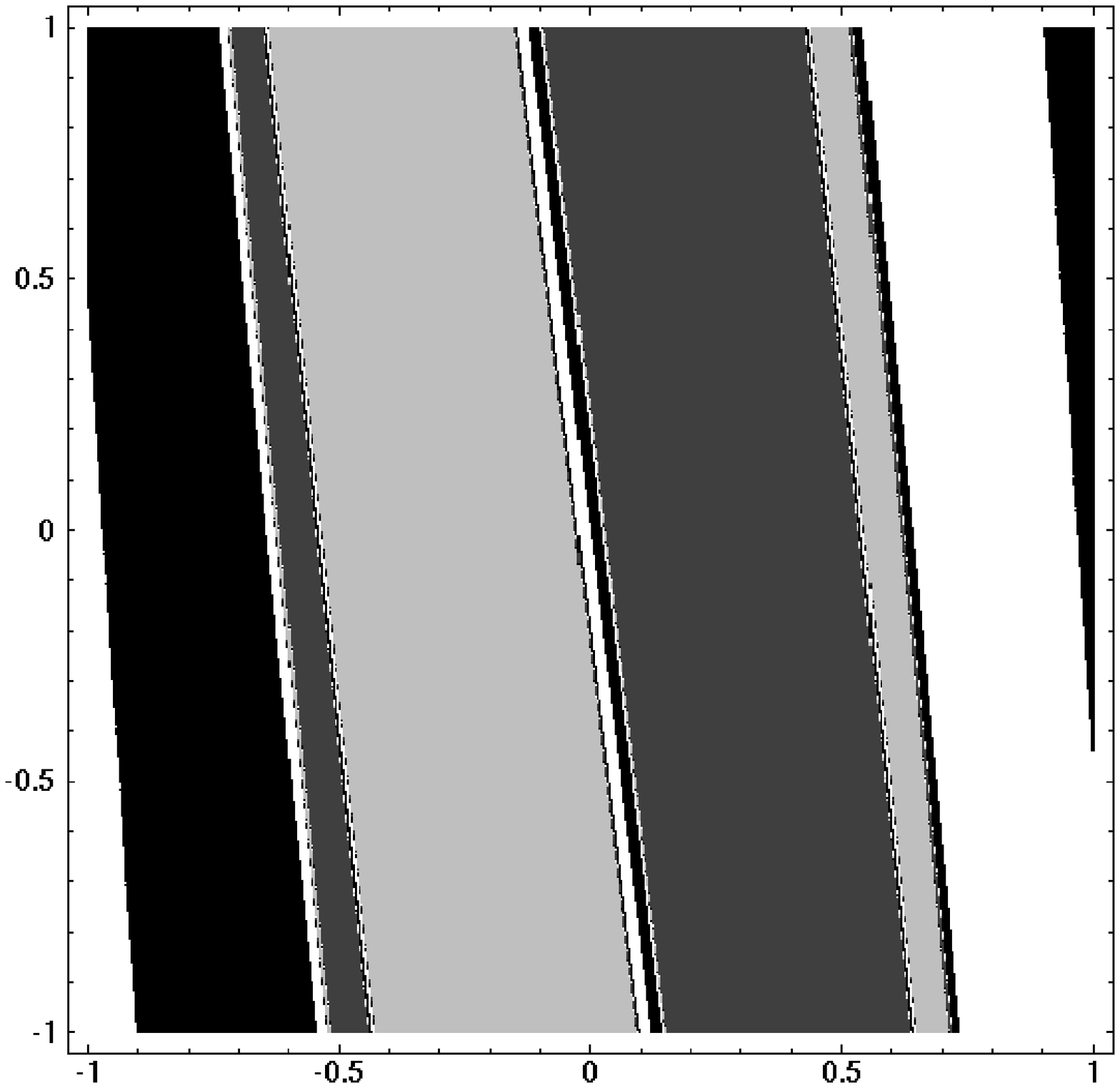}
\includegraphics{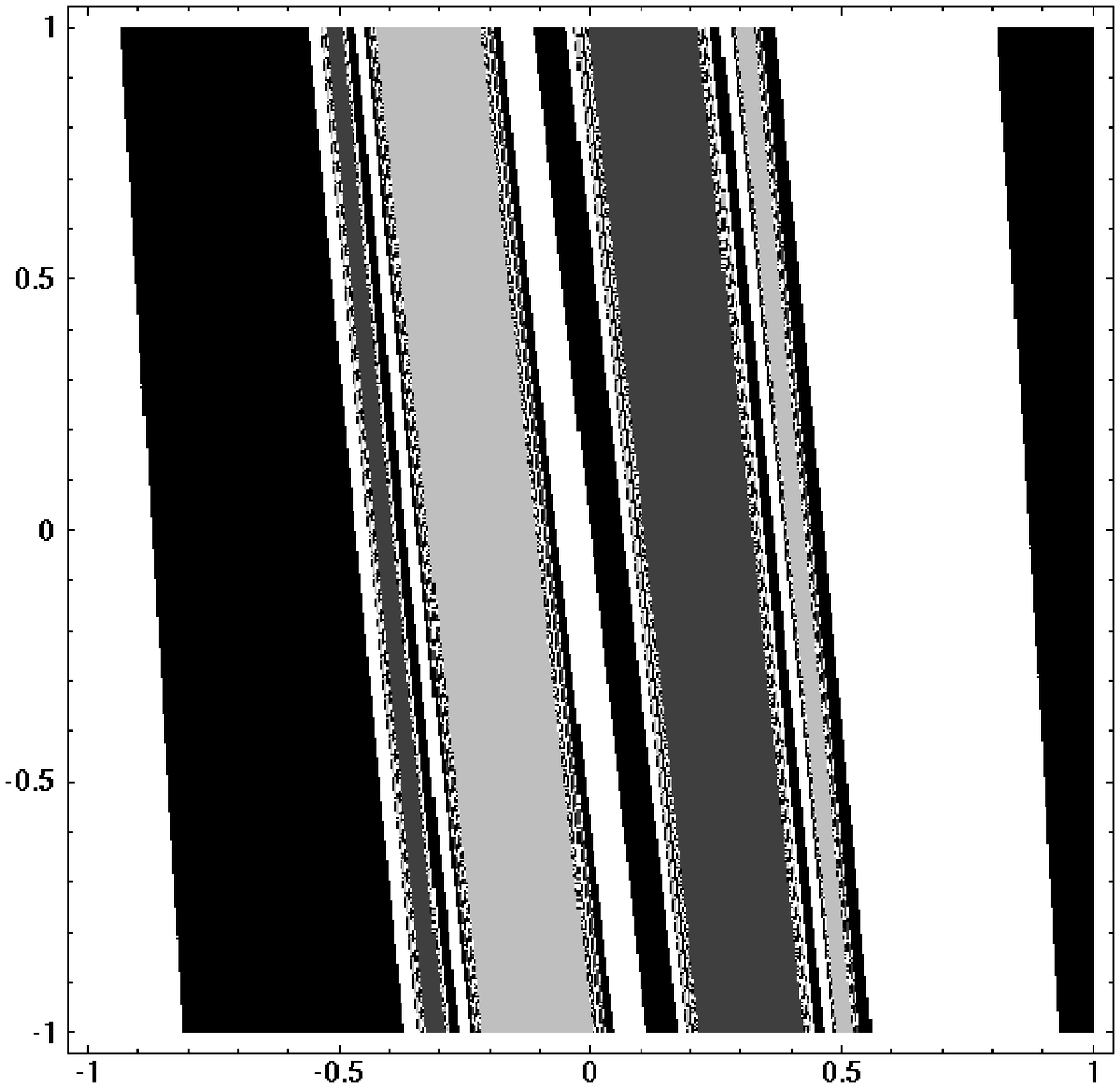}
\includegraphics{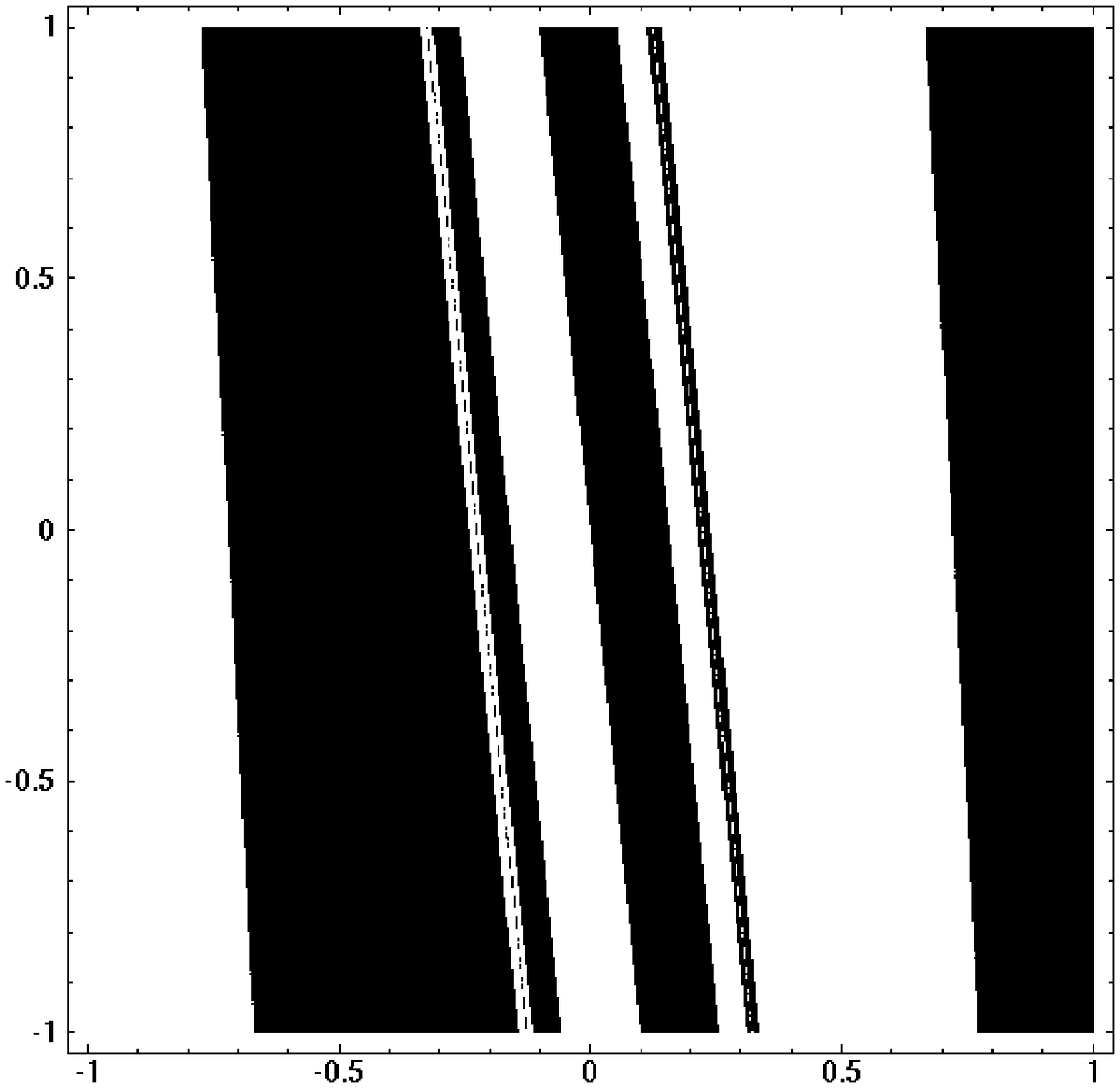}
\vspace{20mm}
\caption{Outcomes for null geodesics in two centre geometries with
$a=0,\; 1$ and $\protect\sqrt{3}$ (arranged from top to bottom).}
\end{figure} 
\vspace*{-30mm}
\begin{picture}(0,0)
\put(110,95){$y_0$}
\put(230,-20){$\theta_0$}
\end{picture}

\newpage

\
\begin{figure}[h]
\vspace{65mm}
\includegraphics{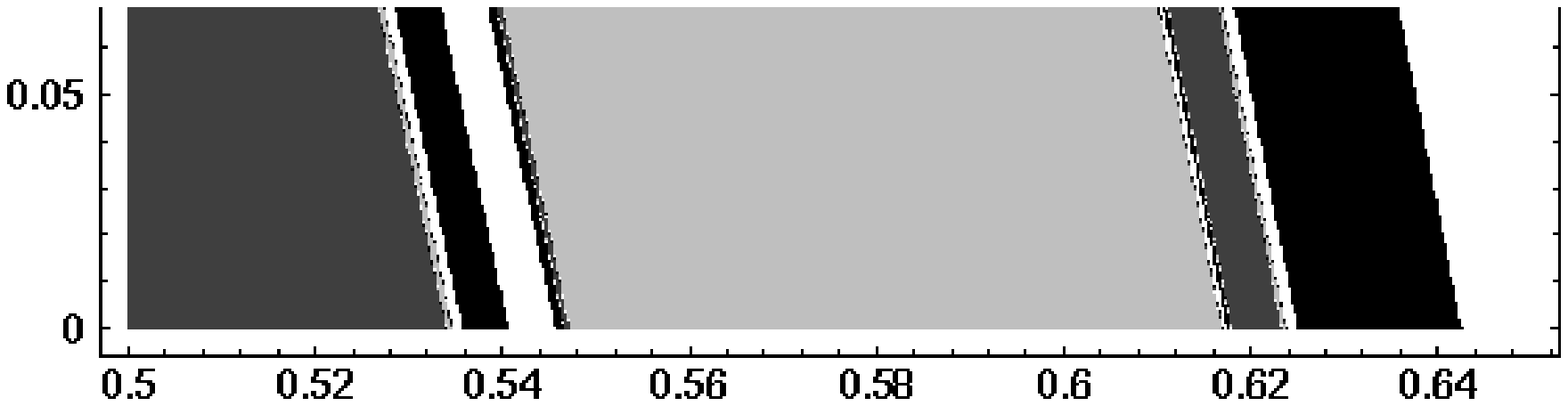}
\includegraphics{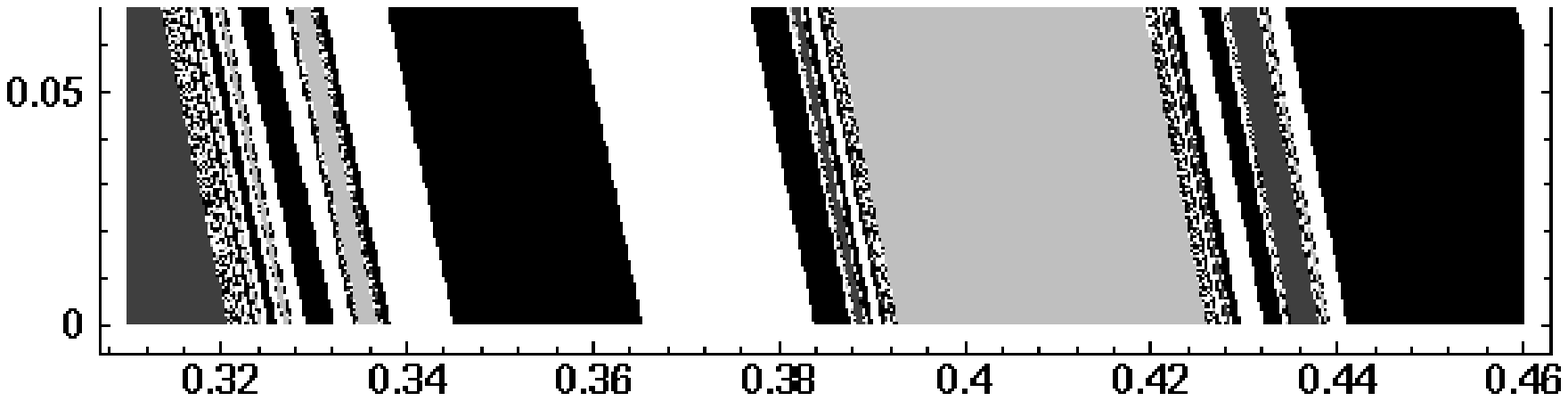}
\includegraphics{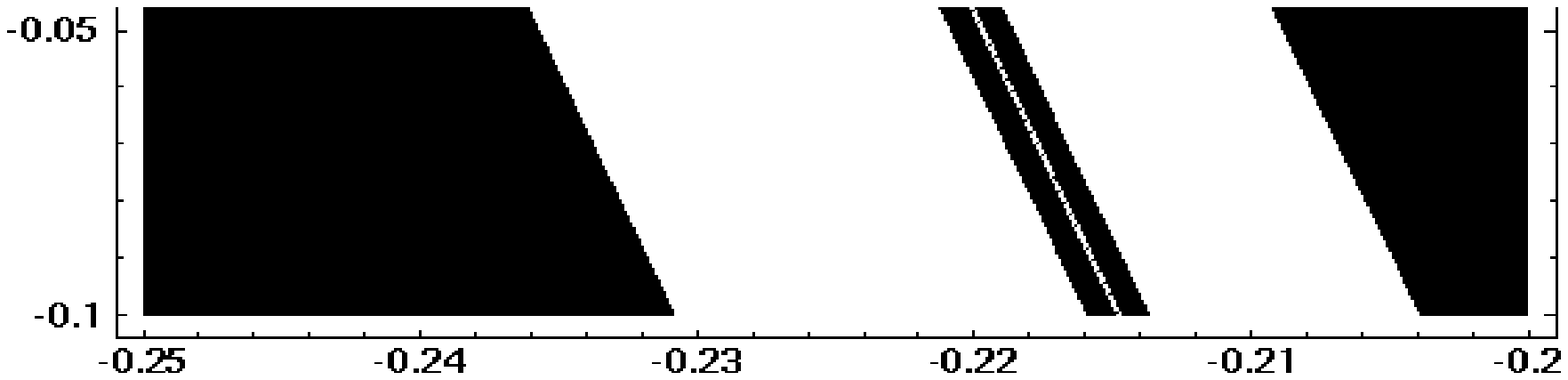}
\vspace{10mm}
\caption{Details of Fig.~2}
\end{figure} 
\begin{picture}(0,0)
\put(90,160){$y_0$}
\put(210,50){$\theta_0$}
\end{picture}

\vspace*{-5mm}

\
\begin{figure}[h]\label{orbs}
\vspace*{75mm}
\includegraphics{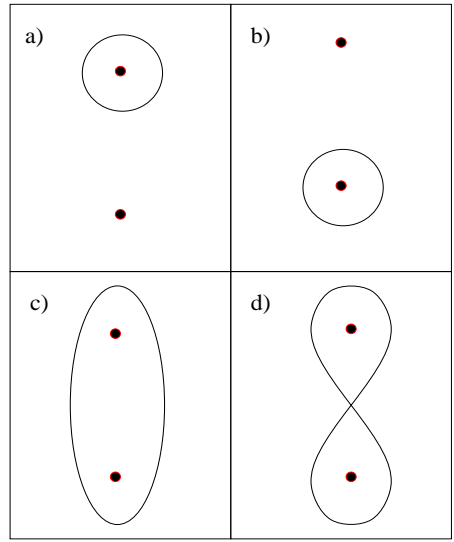}
\vspace*{5mm}
\caption{Primary unstable orbits.}
\end{figure}

\
\begin{figure}[h]\label{symbols}
\vspace*{60mm}
\includegraphics{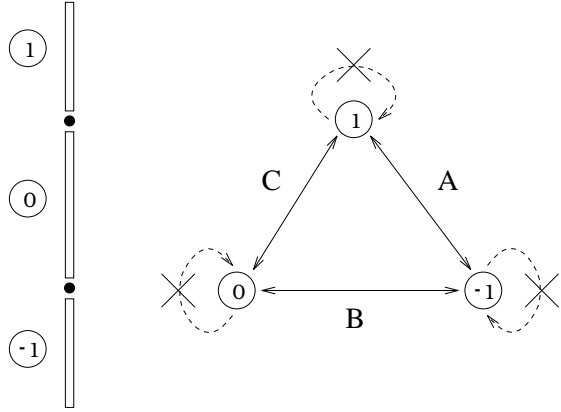}
\caption{The symbolic dynamics.}
\end{figure}

\
\begin{figure}[h]
\vspace{80mm}
\includegraphics{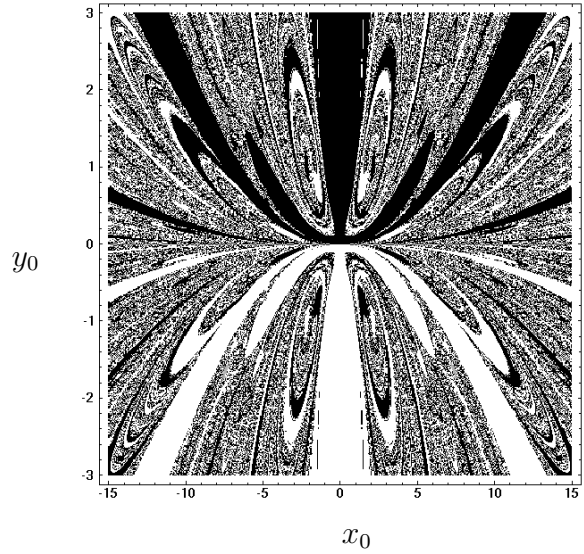}
\vspace{20mm}
\caption{Outcomes for massive, uncharged test particles released from
rest in the $(a=\protect\sqrt{3})$ two centre geometry.}
\end{figure}

\vspace*{-6mm}
\begin{picture}(0,0)
\put(90,180){$y_0$}
\put(215,75){$x_0$}
\end{picture}

\vspace*{2mm}

\newpage
\
\begin{figure}[h]
\vspace{80mm}
\includegraphics{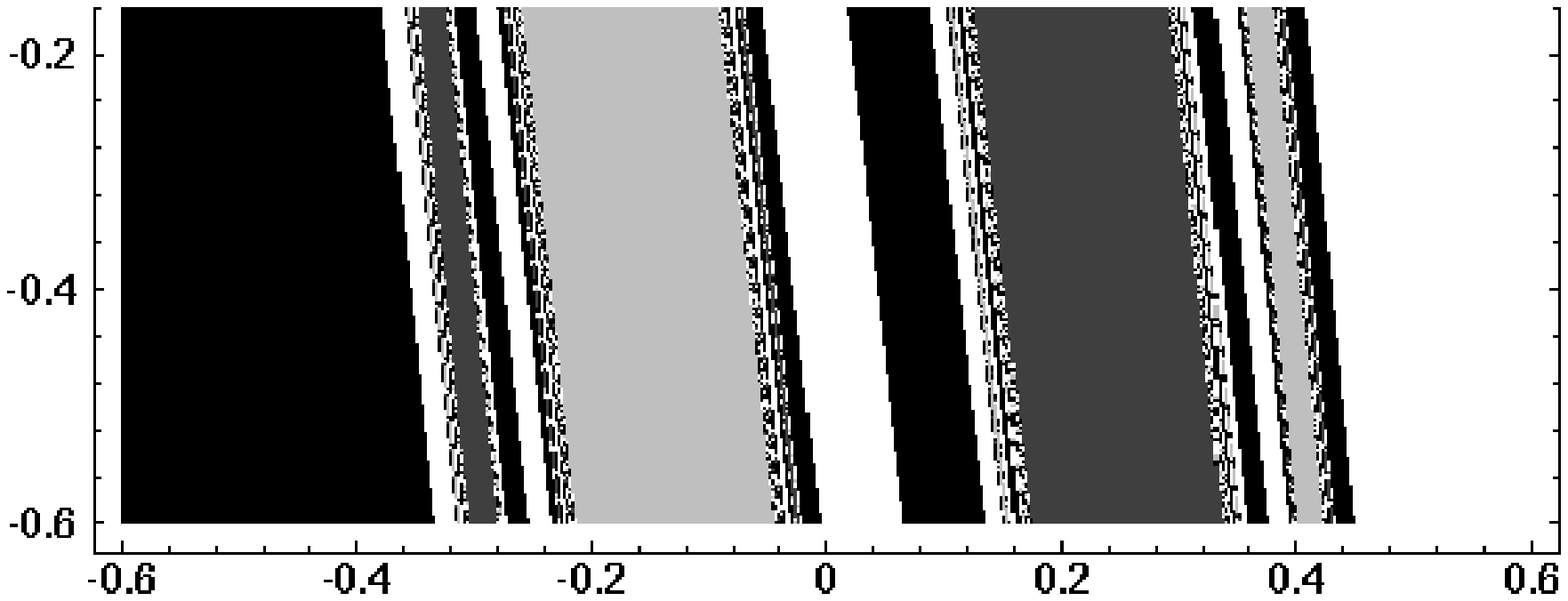}
\includegraphics{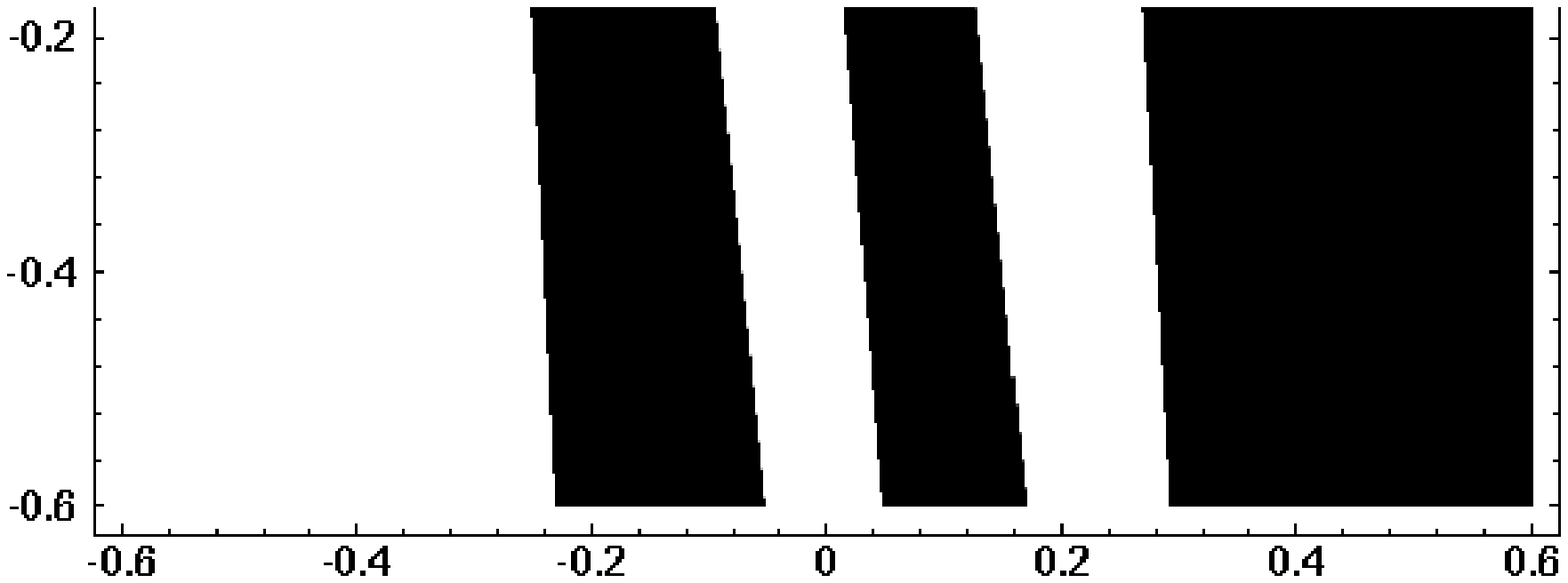}
\vspace{20mm}
\caption{Outcome basins for extremal test particles in geometries with
$a=0$ and $a=\protect\sqrt{3}$ (arrange top to bottom).}
\end{figure} 
\begin{picture}(0,0)
\put(80,200){$y_0$}
\put(210,95){$\theta_0$}
\end{picture}

\vspace*{-5mm}

\end{document}